\documentclass{sf2a-conf2014}
\usepackage{graphicx}
\usepackage{hyperref}
\usepackage[]{natbib}  
\usepackage{epstopdf}

\def\BibTeX{{\rm B\kern-.05em{\sc i\kern-.025em b}\kern-.08em
    T\kern-.1667em\lower.7ex\hbox{E}\kern-.125emX}}
\bibpunct{(}{)}{;}{a}{}{,}  

\newcommand{\meta}{\hbox{[M/H]}}
\newcommand{\feh}{\hbox{[Fe/H]}}

\newcommand{\vphi}{V_\phi}
\def\kms{\,{\rm km\,s^{-1}}}
\def\kpc{\,{\rm kpc}}

\def\dex{\,{\rm dex}}

\def\mag{\,{\rm mag}}

\begin{document}

\TitreGlobal{SF2A 2014}


\title{RAVE: results and updates from Data Release 4}

\runningtitle{RAVE: results and updates from Data Release 4}

\author{G. Kordopatis$^{1,}$}\address{Leibniz-Institut f\"ur  Astrophysik Potsdam (AIP), An der Sternwarte 16, 14482 Potsdam, Germany}\address{Institute of Astronomy, University of Cambridge, Madingley Road, Cambridge CB3 0HA, UK}\

\author{RAVE collaboration}




\setcounter{page}{237}


\maketitle


\begin{abstract}
The \emph{RAdial Velocity Experiment} (RAVE) published in November 2013 its fourth data release with the stellar atmospheric parameters, abundances, distances, radial velocities, proper motions and spectral morphological flags for more than $4\times 10^5$ targets. With that, a plethora of papers ranging from the mass of the Milky Way to the mapping of the Diffuse Interstellar Band, and from the chemo-dynamical history and properties of the disc to the Galaxy's bar pattern speed have also been published, therefore preparing the ground for the exploitation of the Gaia catalogs, since RAVE will be the biggest available spectroscopic database in the magnitude range of Gaia. Here, we review some of these results and present some perspectives about the future data releases.
\end{abstract}



\section{Introduction}
The \emph{Radial Velocity Experiment} \citep{Steinmetz06} is to this date the largest spectroscopic survey of Milky Way stars available for the community. The project finished its observations in April 2013, obtaining more than half a million spectra of more than 400,000 stars in the magnitude range $8<I<12\mag$. 
 RAVE used the 6dF instrument mounted on the Schmidt telescope of the AAO in Siding Spring, Australia. The targeted spectral region (8410--8794\AA), 
 contains the infrared Calcium triplet and is similar to the wavelength chosen for Gaia's \emph{Radial Velocity Spectrometer} \citep[RVS,][]{Cropper11}. The effective resolution of $R\sim7000$ enables us to measure line-of-sight velocities with a median precision better than $1.5\kms$. The distribution on the sky of the observed RAVE targets, as of the end of the project (April 2013), is shown in Fig.~\ref{fig:RAVE_coordinates} and covers a large fraction of the sky accessible from the southern hemisphere.
One of the major improvements of RAVE  fourth data release  \citep[noted DR4 hereafter,][]{Kordopatis13b}, compared to
the previous data releases \citep{Zwitter08,Siebert11}, is its more thorough
metallicity calibration, based on the RAVE observations of cluster stars and
the availability of high-resolution spectra of already observed RAVE targets. In this proceeding we therefore focus in summarising the results of RAVE obtained with DR4. Results of the project up to DR3 can be found summarised in \citet{Siebert12}. Section~\ref{sect:DR4_presentation} presents the new algorithms used to obtain the stellar parameters, abundances and distances. Section~\ref{sect:results_dynamics} and Sect.~\ref{sect:results_chemistry} present  the main results obtained on the Milky Way structure, dynamics and evolution. Finally, Sect.~\ref{sect:conclusions} presents future prospects of the RAVE project. 
 

\section{New pipelines used in the fourth data-release}
\label{sect:DR4_presentation}
\subsection{Obtention of the atmospheric parameters, chemical abundances and detection of chromospheric activity}
A new pipeline for spectral automatic parameterisation has been applied on the totality of the RAVE spectra, based on the one presented in \citet{Kordopatis11a}. Compared to previous data-releases, the spectral degeneracies are better taken into account, since the synthetic grid that has been used has a reduced dimentionality of the parameter space, excludes the core of the Calcium triplet lines and takes into account photometric constraints from 2MASS.  The effective temperatures, surface gravities and overall metallicities are hence computed in that frame. In addition, the chemical abundance pipeline presented in \citet{Boeche11} has been improved, in order to obtain the elemental abundances of six elements, namely the magnesium, aluminium, silicon, titanium, iron, and nickel. 
In the same way as for the previous data releases, the methods presented in \citet{Matijevic11,Matijevic12,Zerjal13} identify and flag, based on the spectral morphology, the spectroscopic binaries and peculiar stars, such for example the ones with chromospheric activity. All the above pipelines allow to obtain all the possible parameters relative to the stellar atmospheres, that will be used as input for the computation of the stellar distances and from there the measurement of the global Galactic chemo-dynamical properties.  

\subsection{Computation of the line-of-sight distances}
\begin{figure}[t!]
 \centering
 \includegraphics[width=0.4\textwidth,height=5.5cm,clip]{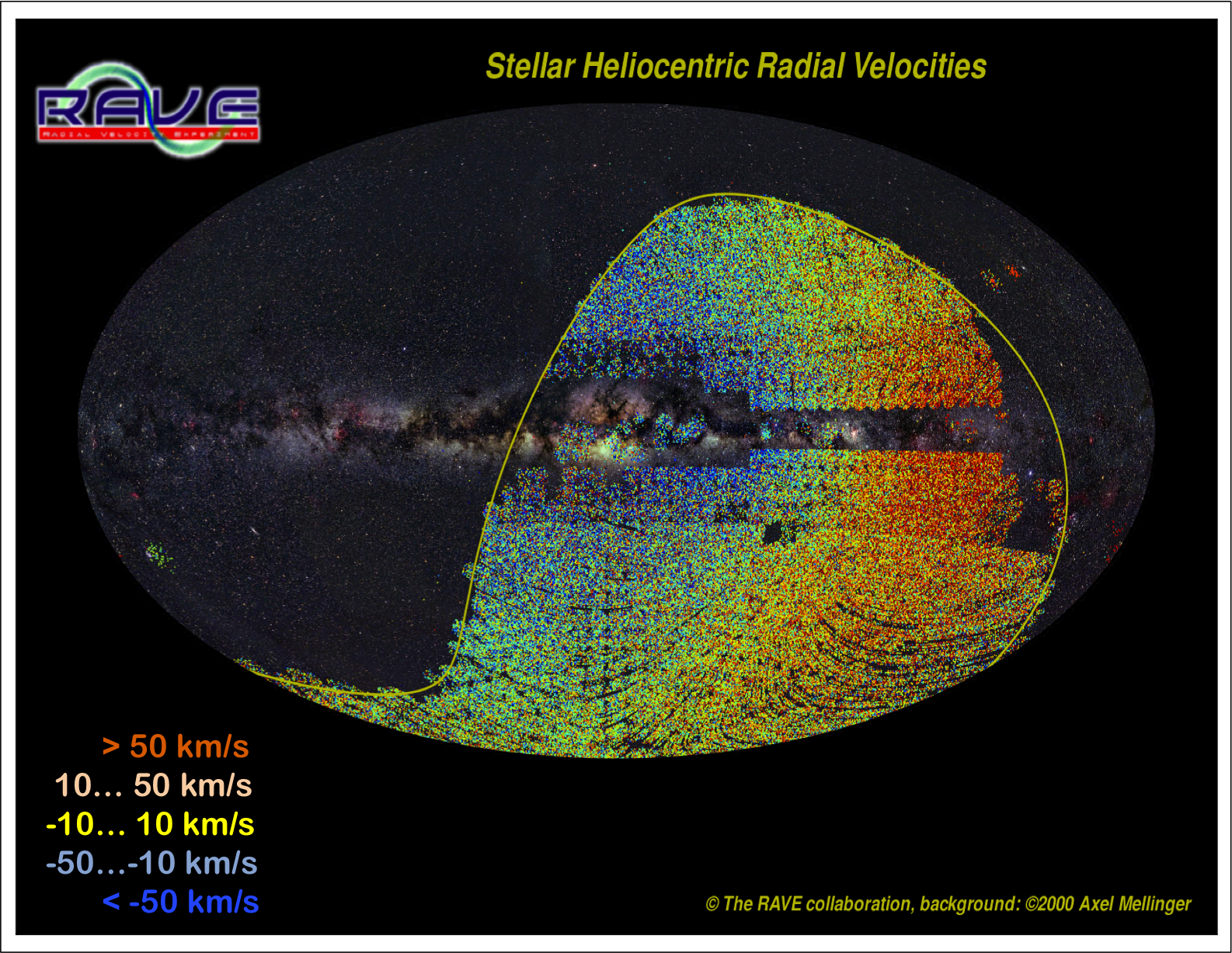}%
 \includegraphics[width=0.4\textwidth,clip]{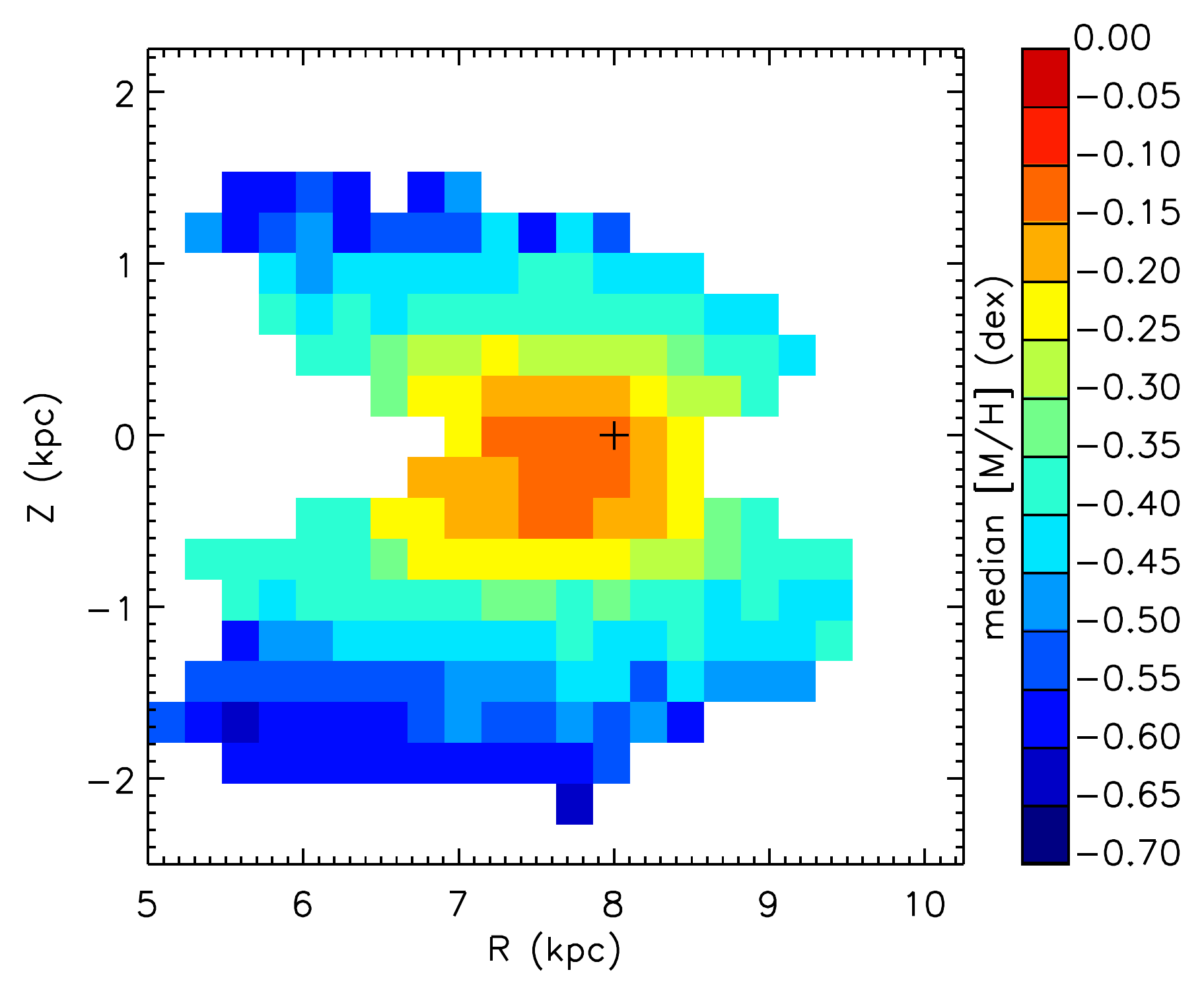}      
  \caption{{\bf Left:} RAVE footprint on the sky as in April 2013, in Galactic coordinates. The points are colour-coded according to their measured heliocentric radial velocity.   {\bf Right:} Median metallicities of the stars as a function of the Galactocentric position \citep[taken from][]{Kordopatis13c}. Assumed Sun's position is indicated by a ``+'' sign, at $(R,Z)=(8,0)\kpc$.}
  \label{fig:RAVE_coordinates}
\end{figure}

RAVE DR4 publishes two different sets of distances, based on different algorithms. The first one, is based on the \citet{Zwitter10} method, that projects the atmospheric parameters on theoretical isochrones introducing only a mild prior on the evolutionary stage at which the star is expected to belong to. The other method, based on the Bayesian approach of \citet{Burnett11} assumes a Galactic model, with realistic stellar and velocity distributions for the Galactic discs and halo, to infer the most likely distance modulus, parallax and line-of-sight distance of the targets \citep{Binney14a}. These distances have been tested on Hipparcos targets \citep{Binney14a} and cluster members \citep[][Anguiano et al. submitted]{Binney14a}, showing relatively small deviations from true values for the types of stars most commonly observed by RAVE. 
The probed distances, colour-coded according to the mean metallicity of the stars are illustrated in Fig.~\ref{fig:RAVE_coordinates} \citep[taken from ][]{Kordopatis13c}.

\section{Milky Way structure and dynamics}
\label{sect:results_dynamics}
Galactic archaeology and the development of models allowing to disentangle to mechanisms responsible for the evolution of our Galaxy rely on our knowledge of the Milky Way's structure and kinematics of its stars. 
\citet{Sharma14} took advantage of the large size of RAVE to measure and constrain the age-velocity dispersion relation for the three kinematic components,  the radial dependence of the velocity dispersions, the solar peculiar motion, the circular speed $\Theta_0$ at the Sun, and the fall of mean azimuthal motion with height above the mid-plane. They found that the Shu distribution describes the best the kinematic distributions of the stars and that the radial scale length of the velocity dispersion profile of the thick disc is smaller than that of the thin disc. 
\citet{Binney14b} also highlights the non-gaussianity of the velocity distribution functions and give formulae from which the shape and orientation of the velocity ellipsoid can be determined at any location.

On the other hand, \citet{Williams13} studied in detail the stellar kinematics in the solar suburb, concentrating on north-south differences. They find a complex three-dimensional structure of velocity space, with among others a clear vertical rarefaction-compression pattern up to $2\kpc$ above the plane, suggestive of wave-like behaviour produced by either internal evolution of the Galaxy  or external factors such as accretions.

Using a statistical sample ten times larger than the one previously used by \citet{Smith07} in DR1, the Galactic escape speed has been re-evaluated by \citet{Piffl14a} to $533^{+54}_{-41}\kms$, confirming previous results and constraining even more the 90\% confidence interval. From the escape speed the authors further derived estimates of the mass of the Galaxy using a simple mass model and found that the dark matter and baryon mass interior to $\rm R_{340}$ is $1.3^{+0.4}_{-0.3} \times 10^{12} \rm M_\odot$, in good 
agreement with recently published mass estimates based on the kinematics of more distant halo stars and the satellite galaxy Leo~I.

\citet{Bienayme14} used a subsample of $\sim 4500$  red clump stars, and determined the vertical force at two distances from the plane, as well as the local dark matter density $\rho_{\rm DM}(z=0)=0.0143\pm0.0011\,{\rm M_\odot\, pc^{-3}}$ and the baryonic surface mass density $\Sigma_{\rm baryons}=44.4\pm4.1\,{\rm M_\odot\,pc^{-2}}$. They found evidence for an unexpectedly large amount of dark matter at distances greater than $2\kpc$ from the plane. 
On the same topic, \citet{Piffl14b} modelled the kinematics of giant stars that lie within $\sim 1.5\kpc$ from the Sun and found that the dark mass contained within the isodensity surface of the dark halo that passes through the Sun ($6\pm0.9\times10^{10}\,\mathrm{M_\odot}$), and the surface density within $0.9\kpc$ of the plane ($69\pm10\,\mathrm{M_\odot\,pc^{-2}}$) are almost independent of the halo's axis ratio $q$. They estimated that the baryonic mass is at most 4.3 per cent of the total Galaxy mass.

Finally, \citet{Antoja14} utilised the moving groups available in the database and found that the azimuthal velocity of the Hercules structure decreases as a function of Galactocentric radius. The authors then modelled this behaviour to impose constraints on the bar's pattern speed. The combined likelihood function of the bar's pattern speed and angle has its maximum for a pattern speed of $\Omega_b = (1.89 \pm 0.08) \times \Omega_0$, where $\Omega_0$ is the local circular frequency. Assuming a Solar radius of $8.05\kpc$ and a local circular velocity of $238\kms$, this corresponds to $\Omega_b = 56 \pm 2\kms\kpc^{-1}$.

\section{Milky Way internal evolution and accretion history}
\label{sect:results_chemistry}
The change in chemical properties of the stars and the inter-stellar medium as a function of position in the Galaxy and the correlation between the stellar kinematics and their atmospheric abundances hold important information on the formation and evolution of the Galaxy's structures.  Their measurement require large statistical samples spanning volumes of several kiloparsecs wide in order to detect trends and identify rare stellar populations.

An estimation of the amount of interstellar matter in the line-of-sight towards the observed stars can be obtained by analysing the absorption lines in the RAVE spectra, originated from the Diffuse Interstellar Bands (DIBs). \citet{Kos13,Kos14} measured the equivalent width of the DIBs present in the spectra, and produced the first pseudo three-dimensional map of the strength of the DIB at 8620\AA, covering the nearest $3\kpc$ from the Sun. The authors have found that the DIB follows the spatial distribution of extinction by interstellar dust, however with a significantly larger vertical scale height. 

Using the fact that DR4 has large statistical sample of stars at different Galactic regions, \citet{Boeche13b,Boeche14} measured the radial and vertical gradients in metallicity and individual $\alpha-$elements in the Galaxy.  They found a radial gradient $\partial \feh / \partial R \sim-0.054\dex\kpc^{-1}$ close to the Galactic plane ($|Z|< 0.4\kpc$) that becomes flatter for larger distances above the plane. Other elements are found to follow the same trend although with some variations from element to element, showing that the thick disc experienced a different chemical enrichment history than the thin disc.

\citet{Kordopatis13c} selected stars located between 1 and $2\kpc$ from the Galactic plane in order to investigate the properties of the metal-weak tail of the thick disc. The authors found a kinematic signature of thick disc stars down to metallicities of $-2\dex$, having the suggested correlation between metallicity and azimuthal velocity of the canonical thick disc stars, of $\partial \vphi / \partial \meta \approx -50 \kms\dex$ \citep[e.g.][]{Kordopatis11b, Kordopatis13a,Lee11}. The authors interpreted this result as evidence that radial migration could not have been the main mechanism at the origin of the formation of the thick disc. 

 \citet{Minchev14} focused on stars in the Solar vicinity, and analysed the velocity dispersion of the giant disc stars as a function of metallicity and $\alpha-$abundances. The authors found that the velocity dispersion of the metal-poor stars first show an increase in their velocity dispersion, then a decrease for the most $\alpha-$enhanced stars ($\rm [Mg/Fe] > 0.4\dex$). By comparing with their chemo-dynamical evolution models of the Milky Way, the authors suggested that this behaviour was evidence of the merger history of the Galaxy and that the stars responsible for the velocity dispersion decrease are stars that reached the Solar neighbourhood through radial migration from the inner Galaxy. Evidence of radial migration towards the metal-rich end distribution in the thin disc has also been found in \citet{Kordopatis14}, based on the orbits of the super-Solar metallicity stars.

Finally, \citet{Kunder14} investigated the chemo-dynamical properties of the stars in the RAVE database around the globular clusters M\,22, NGC\,1851 and NGC\,3201, to assess whether the brightest clusters in the Galaxy might actually be the remnant nuclei of accreted dwarf spheroidal galaxies. The authors report some stars belonging to these clusters being at projected distances of $\sim10$~degrees away from the core of these clusters. In addition, in both of the radial velocity histograms of the regions surrounding NGC\,1851 and NGC\,3201, a peak of stars at $230\kms$ is seen, consistent with extended tidal debris from $\omega$\,Centauri.

\section{Perspectives and relation with Gaia}
\label{sect:conclusions}
Future RAVE data releases are planning, among others, to  improve the calibration of the metal-rich end of the metallicity distribution, thanks to the constant addition of metallicity measurements coming from benchmark stars and high-resolution spectra of super-solar metallicity stars (see Kordopatis et al. 2014, submitted). 
In addition, high-precision APASS photometry will also be included in the pipeline to put more weight on the photometric priors \citep{Munari14}, and hence reducing even more the spectral degeneracies intrinsic to the Calcium triplet wavelength region, responsible to a high extent for the uncertainties in the stellar parameters. Finally, the spectroscopic distances in the catalogue will also be updated for the most metal-poor stars ($\meta<1\dex$), by including lower-metallicity isochrones.

The Gaia satellite will obtain proper motions of exquisite quality and distances derived by parallax, overtaking RAVE's precisions by orders of magnitude \citep{Prusti12}. However, atmospheric parameters and radial velocities for the Gaia stars obtained by the onboard instruments ``Blue'' and ``Red spectro-photometers'' and the RVS will not be published before the end of 2016 \citep{Brown12}. Until then, RAVE will represent the most significant sample in Gaia's magnitude range, from which Galactic archaeology will be possible.

\begin{acknowledgements}
G.K. would like o thank the AS-Gaia and the SF2A for financial support. 
 Funding for RAVE has been provided by: the Australian
Astronomical Observatory; the Leibniz-Institut fuer Astrophysik Potsdam (AIP); the Australian National University;
the Australian Research Council; the French National Research Agency; the German Research Foundation (SPP 1177
and SFB 881); the European Research Council (ERC-StG
240271 Galactica); the Instituto Nazionale di Astrofisica at
Padova; The Johns Hopkins University; the National Science Foundation of the USA (AST-0908326); the W. M.~Keck foundation; the Macquarie University; the Netherlands Research School for Astronomy; the Natural Sciences
and Engineering Research Council of Canada; the Slovenian Research Agency; the Swiss National Science Foundation; the Science \& Technology Facilities Council of the UK;
Opticon; Strasbourg Observatory; and the Universities of
Groningen, Heidelberg and Sydney. 
The research leading to these results has received funding from the European Research
Council under the European Union's Seventh Framework Programme (FP7/2007-2013)/ERC
grant agreement no.\ 321067.
The RAVE web site is
at \url{http://www.rave-survey.org}.
\end{acknowledgements}

\bibliographystyle{aa}  
\bibliography{Kordopatis_SF2A} 

\end{document}